\documentclass[english,twocolumn,twoside]{IEEEtran2e}

\usepackage{amsmath}   

\usepackage{amssymb}   
\usepackage{amscd}
\usepackage{babel}
\usepackage{latexsym}  
\usepackage{epsfig}
\usepackage{bbm}

\newtheorem{defi}{Definition}
\newtheorem{lemma}[defi]{Lemma}
\newtheorem{satz}[defi]{Theorem}
\newtheorem{cor}[defi]{Corollary}
\newtheorem{bem}[defi]{Remark}

\newtheorem{exempel}[defi]{Example}

\newtheorem{conj}[defi]{Conjecture}

\newcommand{\qed}{\hfill $\blacksquare$}
\newcommand{\tr}{{\operatorname{Tr}}}

\newcommand{\bra}[1]{{\langle{#1}|}}
\newcommand{\ket}[1]{{|{#1}\rangle}}

\newcommand{\C}{{\mathbb{C}}}
\newcommand{\R}{{\mathbb{R}}}

\newcommand{\alg}[1]{{\mathfrak{#1}}}
\newcommand{\fset}[1]{{\mathcal{#1}}}

\newcommand{\1}{{\mathbbm{1}}}

\newcommand{\E}{{\operatorname{\mathbb{E}}}}
\newcommand{\Var}{{\operatorname{Var}}}

\begin{document}

\title{Strong Converse for\\ Identification via Quantum Channels}
\author{Rudolf Ahlswede$^*$\thanks{$^*$Fakult\"at f\"ur Mathematik,
Universit\"at Bielefeld, Postfach 100131, 33501 Bielefeld, Germany.} and
Andreas Winter$^\bullet$\thanks{$^\bullet$Department of Computer Science, University of Bristol,
Merchant Venturers Building, Woodland Road, Bristol BS8 1UB, United Kingdom.
Email:~\texttt{winter@cs.bris.ac.uk}.}
\thanks{Version of October 19, 2001.}
}

\maketitle


\begin{abstract}
  In this paper we present a simple proof of the strong converse for
  identification via discrete memoryless quantum channels, based on
  a novel covering lemma. The new method is
  a generalization to quantum communication channels of Ahlswede's recently
  discovered appoach to classical channels. It
  involves a development of explicit large deviation
  estimates to the case of random variables taking values in selfadjoint
  operators on a Hilbert space. This theory is presented separately
  in an appendix, and we illustrate it by showing its application
  to quantum generalizations of classical hypergraph covering
  problems.
\end{abstract}


\section{Introduction}
\label{sec:intro}
Ahlswede and Dueck~\cite{ahlswede:dueck:1} found the identification
capacity of a discrete memoryless channel by establishing the optimal
(second order) rate via a so--called \emph{soft converse}. Subsequently,
the \emph{strong converse}, conjectured by them, was proved by
Han and Verd\'{u}~\cite{han:verdu:1}. Even their
second, simplified proof~\cite{han:verdu:2} uses rather involved
arguments.
\par
In~\cite{ahlswede:IDconverse} it is shown
how simple ideas regarding coverings of hypergraphs (formalized in
lemma~\ref{lemma:covering:2}) can be used to obtain the approximations of
output statistics needed in the converse.
\par
Formally, we investigate the following situation: consider a discrete
memoryless channel $W^n:{\cal X}^n\rightarrow{\cal Y}^n$ ($n\geq 1$),
i.e.~for $x^n=x_1\ldots x_n\in{\cal X}^n$, $y^n=y_1\ldots y_n\in{\cal Y}^n$
$$W^n(y^n|x^n)=W(y_1|x_1)\cdots W(y_n|x_n),$$
with a channel $W:{\cal X}\rightarrow{\cal Y}$ which we identify with the
DMC. It is well known~\cite{shannon:info} that the transmission
capacity if this channel (with the strong converse proven
by Wolfowitz~\cite{wolfowitz:strong}) is
$$C(W)=\max_{P\text{ p.d. on }{\cal X}} I(P;W).$$
Here $I(P;W)=H(PW)-H(W|P)$ is Shannon's mutual information, where
$PW=\sum_{x\in{\cal X}} P(x)W(\cdot|x)$ is the output distribution
on ${\cal Y}$, and $H(W|P)=\sum_{x\in{\cal X}} P(x)H(W(\cdot|x))$
is the conditional entropy of the channel for the input distribution
$P$.
\par
Ahlwede and Dueck, considering not the problem that the receiver wants to
recover a message (\emph{transmission problem}), but wants to decide
whether or not the sent message is identical to an arbitrarily chosen
one (\emph{identification problem}), defined a
$(n,N,\lambda_1,\lambda_2)$ identification (ID) code to be a collection of
pairs
$$\{(P_i,{\cal D}_i):i=1,\ldots,N\},$$
with probability distributions $P_i$ on ${\cal X}^n$ and
${\cal D}_i\subset{\cal Y}^n$, such that the error probabilities
of first resp. second kind satisfy
\begin{align*}
  P_iW^n({\cal D}_i^c)
             &=\sum_{x^n\in{\cal X}^n} P_i(x^n)W^n({\cal D}_i^c|x^n)\leq\lambda_1, \\
  P_jW^n({\cal D}_i)
             &=\sum_{x^n\in{\cal X}^n} P_j(x^n)W^n({\cal D}_i|x^n)\leq\lambda_2,
\end{align*}
for all $i,j=1,\ldots,N$, $i\neq j$. Here ${\cal D}_i^c={\cal Y}^n\setminus{\cal D}_i$
is the set complement of ${\cal D}_i$ in ${\cal Y}^n$, and
$$W^n({\cal A}|x^n)=\sum_{y^n\in{\cal A}} W^n(y^n|x^n)$$
is a convenient shortcut for the probability of an event
${\cal A}\subset{\cal Y}^n$ conditional on $x^n$.
Define $N(n,\lambda_1,\lambda_2)$ to be the maximal $N$ such that
a $(n,N,\lambda_1,\lambda_2)$ ID code exists.
\par
With these definitions one has
\begin{satz}[Ahlswede, Dueck~\cite{ahlswede:dueck:1}]
  \label{satz:ID:coding}
  For every $\lambda_1,\lambda_2>0$ and $\delta>0$, and for every
  sufficiently large $n$
  $$N(n,\lambda_1,\lambda_2)\geq \exp(\exp(n(C(W)-\delta))).$$
  \qed
\end{satz}
The work~\cite{ahlswede:IDconverse} is devoted to a comparably short,
and conceptually simple proof of
\begin{satz}
  \label{satz:ID:converse}
  Let $\lambda_1,\lambda_2>0$ such that $\lambda_1+\lambda_2<1$. Then
  for every $\delta>0$ and every sufficiently large $n$
  $$N(n,\lambda_1,\lambda_2)\leq \exp(\exp(n(C(W)+\delta))).$$
  \qed
\end{satz}
Note that for $\lambda_1+\lambda_2\geq 1$ no upper bound on
$N(n,\lambda_1,\lambda_2)$ can hold:
a successful strategy would be that the receiver ignores the actual signal, and
to identify $i$ guesses YES with probability $1-\lambda_1$, NO with probability
$\lambda_1\geq 1-\lambda_2$.
\par
The first proof of theorem~\ref{satz:ID:converse} was given
in~\cite{han:verdu:1}, the method to be further
extended in~\cite{han:verdu:2}. In~\cite{ahlswede:IDconverse} it is
returned to the very first idea
from~\cite{ahlswede:dueck:1}, essentially to replace the distributions
$P_i$ by uniform distributions on ``small'' subsets of ${\cal X}^n$,
namely with cardinality slightly above $\exp(nC(W))$.
\par
L\"ober~\cite{loeber:diss} began the study of identification via quantum
channels. Following his work, and after Holevo~\cite{holevo:bound},
we define a (discrete memoryless) classical--quantum channel (quantum
channel for short) to be a map
$$W:{\cal X}\longrightarrow {\cal S}({\cal H}),$$
with ${\cal X}$ a finite set, as before, and ${\cal S}({\cal H})$ the set of quantum
states of the complex Hilbert space ${\cal H}$, which we assume to be finite dimensional.
In the sequel, we shall use $a=|\fset{X}|$ and $d=\dim{\cal H}$.
We identify ${\cal S}({\cal H})$, as usual, with the set of density operators, i.e.
the selfadjoint, positive semidefinite, linear operators on ${\cal H}$ with unit
trace\footnote{See Davies~\cite{davies:etc} for the mathematics
  to describe quantum systems.}:
$${\cal S}({\cal H})=\{\rho: \rho=\rho^*\geq 0,\tr\rho=1\}.$$
In the sequel we will write $W_x$ for the images $W(x)$
of the channel map.
\par
Associated to $W$ is the channel map on $n$--blocks
$$W^n:{\cal X}^n\longrightarrow {\cal S}({\cal H}^{\otimes n}),$$
with
$$W^n_{x^n}=W_{x_1}\otimes\cdots\otimes W_{x_n}.$$
One can use quantum channels to transmit classical information,
and Holevo~\cite{holevo:qucapacity} showed that the capacity is
$$C(W)=\max_{P\text{ p.d. on }{\cal X}} I(P;W).$$
Here $I(P;W)=H(PW)-H(W|P)$ is the von Neumann mutual information, with
the output state $PW=\sum_{x\in{\cal X}} P(x)W_x$
on ${\cal H}$, and $H(W|P)=\sum_{x\in{\cal X}} P(x)H(W_x)$
the conditional entropy of the channel for the input distribution $P$.
The only difference to Shannon's result is that here
$H$ denotes the \emph{von Neumann} entropy which is defined, for
a state $\rho$, as
$$H(\rho)=-\tr\rho\log\rho.$$
The strong converse for this situation was proved (independently)
in~\cite{ogawa:nagaoka} and~\cite{winter:qstrong}.
\par
Quantum channels are a generalization of classical channels
in the following sense: choose any orthonormal basis
$(e_y:y\in{\cal Y})$ of the $|{\cal Y}|$--dimensional
Hilbert space ${\cal H}$, and define for the
classical channel $W:{\cal X}\rightarrow{\cal Y}$
the corresponding quantum channel
$\widetilde{W}:{\cal X}\rightarrow{\cal S}({\cal H})$ by
$$\widetilde{W}_x=\sum_y W(y|x)\ket{e_y}\bra{e_y}.$$
Obviously for another channel $V$ one has
$\widetilde{V\!\!\times\! W}=\widetilde{V}\otimes\widetilde{W}$.
\par
Regarding the decoding sets let ${\cal D}\subset{\cal Y}$, then
the corresponding operator $D=\sum_{y\in{\cal D}}\ket{e_y}\bra{e_y}$
satisfies for all $x$
$$W({\cal D}|x)=\tr(\widetilde{W}_x D).$$
Observe that by this translation rule a partition of ${\cal Y}$
corresponds to a projection valued measure (PVM) on ${\cal H}$,
i.e.~a collection of mutually orthogonal projectors which sum
to $\1$. Conversely, given any operator $D$ on ${\cal H}$
with $0\leq D\leq\1$, define the function
$\delta:{\cal Y}\rightarrow[0,1]$ by
$\delta(y)=\bra{e_y}D\ket{e_y}$. Then for all $x$
$$\tr(\widetilde{W}_x D)=\sum_{y\in{\cal Y}} \delta(y)W(y|x),$$
which implies that every quantum observation, i.e.~a
positive operator valued measure (POVM), of the states
$\widetilde{W}_x$ can be simulated by a classical randomized decision
rule on ${\cal Y}$. One consequence of this is that
the transmission capacities of $W$ and of $\widetilde{W}$ are equal:
$C(W)=C(\widetilde{W})$. Equally, also the identification capacities
(whose definition in the quantum case is given below) coincide.
For randomization at the decoder cannot improve either minimum
error probability.
\par
Abstractly, just given the states $W_x$, this situation occurs
if they pairwise commute: for then they are simultaneously
diagonalizable, hence the orthonormal basis $(e_y:y\in{\cal Y})$
arises.
\par
According to~\cite{loeber:diss} a $(n,N,\lambda_1,\lambda_2)$
\emph{quantum identification (QID) code}
is a collection of pairs
$$\{(P_i,D_i):i=1,\ldots,N\},$$
with probability distributions $P_i$ on ${\cal X}^n$,
and operators $D_i$ on ${\cal H}^{\otimes n}$ satisfying
$0\leq D_i\leq\1$, such that the error probabilities of
first resp. second kind satisfy
\begin{align*}
   \tr &\left(P_iW^n(\1-D_i)\right) \\
        &=\tr\left(\left(\sum_{x^n\in{\cal X}^n} P_i(x^n)W_{x^n}\right)(\1-D_i)\right)
              \leq\lambda_1, \\
   \tr &\left(P_jW^n\cdot D_i\right) \\
        &=\tr\left(\left(\sum_{x^n\in{\cal X}^n} P_j(x^n)W_{x^n}\right)D_i\right)
              \leq\lambda_2,
\end{align*}
for all $i,j=1,\ldots,N$, $i\neq j$. Again, define
$N(n,\lambda_1,\lambda_2)$ to be the maximal $N$ such that
a $(n,N,\lambda_1,\lambda_2)$ QID code exists.
\par
This definition has a subtle problem: since the $D_i$ need not commute,
it is possible that identifying for a message $i$ prohibits identification
for $j$, as the corresponding POVMs $(D_i,\1-D_i)$ and
$(D_j,\1-D_j)$ may be incompatible. To allow simultaneous
identification of all messages we have to assue that the
$D_i$ have a common refinement, i.e.~there exists
a POVM $(E_k:k=1,\ldots,K)$ and subsets $I_i$ of
$\{1,\ldots,K\}$ such that
$$D_i=\sum_{k\in I_i} E_k,$$
for all $i$. In this case the QID code is called \emph{simultaneous},
and $N_{\text{sim}}(n,\lambda_1,\lambda_2)$ is the maximal $N$ such that
a simultaneous $(n,N,\lambda_1,\lambda_2)$ quantum identification code exists.
Clearly 
$$N_{\text{sim}}(n,\lambda_1,\lambda_2)\leq N(n,\lambda_1,\lambda_2).$$
In analogy to the above theorems it was proved:
\begin{satz}[L\"ober~\cite{loeber:diss}]
  \label{satz:SQID:coding:converse}
  For every $\lambda_1,\lambda_2>0$ and $\delta>0$, and for every
  sufficiently large $n$
  $$N_{\text{sim}}(n,\lambda_1,\lambda_2)\geq \exp(\exp(n(C(W)-\delta))).$$
  On the other hand, let $\lambda_1,\lambda_2>0$
  such that $\lambda_1+\lambda_2<1$. Then
  for every $\delta>0$ and every sufficiently large $n$
  $$N_{\text{sim}}(n,\lambda_1,\lambda_2)\leq \exp(\exp(n(C(W)+\delta))).$$
  \qed
\end{satz}
Looking at the examples given in~\cite{ahlswede:dueck:1}
the simultaneity condition seems completely natural.
But this need not always be the case.
\begin{exempel}
  Modify the ``sailors' wives'' situation (Example
  $1$ form~\cite{ahlswede:dueck:1}) as follows:
  the $N$ sailors are not married each to one wife
  but instead are all in love with a single girl.
  One day in a storm one sailor drowns, and his identity should
  be communicated home. The girl however is capricious
  to the degree that it is impossible to predict who is her
  sweetheart at a given moment: when the message about the drowned
  sailor arrives, she will only ask for her present sweetheart,
  and only she will ask.
\end{exempel}
\par\medskip
With our present approach we can get rid of the simultaneity
condition in the converse (whereas by the above theorem identification
codes approaching the capacity can be designed to be simultaneous
--- namely, by~\cite{ahlswede:dueck:1} for any sort of channel and a transmission
code of rate $R$ for it, one can construct an ID code ``on top'' of the
transmission code, and with identification rate $R$, asymptotically):
\begin{satz}
  \label{satz:QID:converse}
  Let $\lambda_1,\lambda_2>0$ such that $\lambda_1+\lambda_2<1$. Then
  for every $\delta>0$ and every sufficiently large $n$
  $$N(n,\lambda_1,\lambda_2)\leq \exp(\exp(n(C(W)+\delta))).$$
\end{satz}
The rest of the paper is divided into two major
blocks: first, after a short review of the ideas from~\cite{ahlswede:IDconverse}
in section~\ref{sec:ID:proof},
the rest of the main text will be devoted to the proof of
theorem~\ref{satz:QID:converse}
(as explained, this contains theorem~\ref{satz:ID:converse} indeed
as a special case), in section~\ref{sec:QID:proof}.
\par
The other block is the appendix, containing the fundamentals of a theory
of (selfadjoint) operator valued random variables. There the large
deviation bounds to be used in the main text are derived.

\section{The classical case}
\label{sec:ID:proof}
The core of the proof of theorem~\ref{satz:ID:converse}
in~\cite{ahlswede:IDconverse} is the following result about hypergraphs.
Recall that a \emph{hypergraph} is a pair $\Gamma=({\cal V},{\cal E})$
with a finite set ${\cal V}$ of vertices, and a finite set
${\cal E}$ of (hyper--) edges $E\subset{\cal V}$. We call
$\Gamma$ $e$--uniform, if all its edges have cardinality $e$.
For an edge $E\in{\cal E}$ denote the characteristic function of
$E\subset{\cal V}$ by $1_E$.
\par
The starting point is a result from large
deviation theory:
\begin{lemma}
  \label{lemma:sanov}
  For an i.i.d.~sequence $Z_1,\ldots, Z_L$ of random variables
  with values in $[0,1]$ with expectation $\E Z_i=\mu$, and
  $0<\epsilon<1$
  \begin{align*}
    \Pr\left\{ \frac{1}{L}\sum_{i=1}^L Z_i>(1+\epsilon)\mu \right\}
                                     &\leq\exp(-L D((1+\epsilon)\mu\|\mu)), \\
    \Pr\left\{ \frac{1}{L}\sum_{i=1}^L Z_i<(1-\epsilon)\mu \right\}
                                     &\leq\exp(-L D((1-\epsilon)\mu\|\mu)),
  \end{align*}
  where $D(\alpha\|\beta)$ is the information divergence of the binary
  distributions $(\alpha,1-\alpha)$ and $(\beta,1-\beta)$. Since
  for $-\frac{1}{2}\leq x \leq\frac{1}{2}$
  $D((1+x)\mu\|\mu)\geq \frac{1}{2\ln 2}x^2\mu$,
  it follows that
  $$\Pr\left\{ \frac{1}{L}\sum_{i=1}^L Z_i\not\in[(1-\epsilon)\mu,(1+\epsilon)\mu] \right\}
                             \leq 2\exp\left(-L\!\cdot\! \frac{\epsilon^2\mu}{2\ln 2}\right)\!.$$
  \qed
\end{lemma}
\begin{lemma}
  \label{lemma:covering:1}
  Let $\Gamma=({\cal V},{\cal E})$ be an $e$--uniform
  hypergraph, and $P$ a probability distribution on ${\cal E}$.
  Define the probability distribution $Q$ on ${\cal V}$ by
  $$Q(v)=\sum_{E\in{\cal E}} P(E)\frac{1}{e}1_E(v),$$
  and fix $\epsilon,\tau>0$. Then there exist vertices
  ${\cal V}_0\subset{\cal V}$ and edges
  $E_1,\ldots,E_L\in{\cal E}$ such that with
  $$\bar{Q}(v)=\frac{1}{L}\sum_{i=1}^L \frac{1}{e}1_{E_i}(v)$$
  the following holds:
  $$Q({\cal V}_0)\leq\tau,$$
  $$\forall v\in{\cal V}\setminus{\cal V}_0\ \ 
                (1-\epsilon)Q(v)\leq \bar{Q}(v)\leq (1+\epsilon)Q(v),$$
  $$L\leq 1+\frac{|{\cal V}|}{e}\frac{2\ln 2\log(2|{\cal V}|)}{\epsilon^2\tau}.$$
\end{lemma}
\begin{proof}
  See~\cite{ahlswede:IDconverse}.
\end{proof}
For ease of application we formulate a slightly more general
version of this:
\begin{lemma}
  \label{lemma:covering:2}
  Let $\Gamma=({\cal V},{\cal E})$ be a hypergraph, with a
  measure $Q_E$ on each edge $E$, such that
  $Q_E(v)\leq\eta$ for all $E$, $v\in E$.
  For a probability distribution $P$ on ${\cal E}$ define
  $$Q=\sum_{E\in{\cal E}} P(E)Q_E,$$
  and fix $\epsilon,\tau>0$.Then there exist vertices
  ${\cal V}_0\subset{\cal V}$ and edges
  $E_1,\ldots,E_L\in{\cal E}$ such that with
  $$\bar{Q}=\frac{1}{L}\sum_{i=1}^L Q_{E_i}$$
  the following holds:
  $$Q({\cal V}_0)\leq\tau,$$
  $$\forall v\in{\cal V}\setminus{\cal V}_0\ \ 
                (1-\epsilon)Q(v)\leq \bar{Q}(v)\leq (1+\epsilon)Q(v),$$
  $$L\leq 1+\eta|{\cal V}|\frac{2\ln 2\log(2|{\cal V}|)}{\epsilon^2\tau}.$$
  \qed
\end{lemma}
The interpretation of this result is as follows: $Q$ is the expectation
measure of the measures $Q_E$, which are sampled by the $Q_{E_i}$. The
lemma says how close the sampling average $\bar{Q}$ can be to $Q$.
In fact, assuming $Q_E(E)=q\leq 1$ for all $E\in{\cal E}$,
one easily sees that
$$\|Q-\bar{Q}\|_1\leq 2\epsilon+2\tau.$$
\par
The idea for the proof of theorem~\ref{satz:ID:converse} is now:
to replace the (in principle) arbitrary distributions $P_i$
on ${\cal X}^n$ of a $(n,N,\lambda_1,\lambda_2)$ ID code
$\{(P_i,D_i):i=1,\ldots,N\}$, by 
uniform distributions on subsets of ${\cal X}^n$, with
cardinality bounded essentially by $\exp(nC(W))$.
The condition is that the corresponding \emph{output}
distributions are close, so the resulting ID code will
be a bit worse, but still nontrivial.
This is done with the help of the covering lemma~\ref{lemma:covering:2},
applied to typical sequences in ${\cal Y}^n$ as vertices,
and sets of induced typical sequences as edges. For details
see~\cite{ahlswede:IDconverse}.

\section{Proof of theorem~\ref{satz:QID:converse}}
\label{sec:QID:proof}
It was already pointed out in the previous section that the main idea
of the converse proof is to replace the arbitrary code distributions $P_i$
by regularized approximations, the quality of approximation being measured
by the $\|\cdot\|_1$--distance of the output distributions.
\par
Hence, to extend the method to quantum channels we have to use the
$\|\cdot\|_1$--distance of the output \emph{quantum states},
and we have to find quantum versions of the lemmas~\ref{lemma:sanov}
and~\ref{lemma:covering:2}.
\par
Define a \emph{quantum hypergraph} to be a pair $({\cal V},{\cal E})$
with a finite dimensional Hilbert space ${\cal V}$ and a (finite) collection
${\cal E}$ of operators $E$ on ${\cal V}$, $0\leq E\leq\1$ (see
the discussion in subsection~F
of the appendix).
\par
Analogous to lemma~\ref{lemma:sanov} is
theorem~\ref{satz:chernoff} (in the appendix),
the analog of lemma~\ref{lemma:covering:2} is
\begin{lemma}
  \label{lemma:qu:covering}
  Let $({\cal V},{\cal E})$ be a quantum hypergraph such that $E\leq\eta\1$
  for all $E\in{\cal E}$. For a probability distribution $P$ on ${\cal E}$
  define
  $$\rho=\sum_{E\in{\cal E}} P(E)E,$$
  and fix $\epsilon,\tau>0$. Then there exists a subspace
  ${\cal V}_0<{\cal V}$ and $E_1,\ldots,E_L\in{\cal E}$
  such that with
  $$\bar{\rho}=\frac{1}{L}\sum_{i=1}^L E_i,$$
  and denoting by $\Pi_0$ and $\Pi_1$ the orthogonal projections
  onto ${\cal V}_0$ and its complement, respectively, the
  following holds:
  $$\tr\rho\Pi_0\leq\tau,$$
  $$(1-\epsilon)\Pi_1\rho\Pi_1\leq \Pi_1\bar{\rho}\Pi_1
                              \leq (1+\epsilon)\Pi_1\rho\Pi_1,$$
  $$L\leq 1+\eta(\dim{\cal V})\frac{2\ln 2\log(2\dim{\cal V})}{\epsilon^2\tau}.$$
\end{lemma}
\begin{proof}
  Diagonalize $\rho$, i.e.~$\rho=\sum_{j} r_j\pi_j$, and let
  $$\Pi_0=\sum_{j:\ r_j<\tau/\dim{\cal V}} \!\!\pi_j,\quad \Pi_1=\1-\Pi_0.$$
  Clearly $\tr(\rho\Pi_o)\leq\tau$.
  \par
  Now consider the quantum hypergraph
  $({\cal V}_0^\perp,\Pi_1{\cal E}\Pi_1)$,
  whose edges also obey the upper bound $\eta\1$,
  and which has edge average $\Pi_1\rho\Pi_1\geq\tau\1/\dim{\cal V}$.
  \par
  Now let $X_1,\ldots,X_L$ i.i.d.~with $\Pr\{X_i=\Pi_1 E \Pi_1\}=P(E)$.
  Then we can estimate with
  theorem~\ref{satz:chernoff}, applied to the variables
  $\eta^{-1}X_i$:
  \begin{align*}
    \Pr&\left\{ \frac{1}{L}\sum_{i=1}^L X_i\not\in
                                [(1-\epsilon)\Pi_1\rho\Pi_1,(1+\epsilon)\Pi_1\rho\Pi_1] \right\} \\
       &\phantom{========}\leq 2(\dim{\cal V})\exp
                           \left(-L\!\cdot\!\frac{\epsilon^2\tau}{2\eta\dim{\cal V}\ln 2}\right),
  \end{align*}
  which is smaller than $1$ if
  $$L>\eta(\dim{\cal V})\frac{2\ln 2\log(2\dim{\cal V})}{\epsilon^2\tau},$$
  in which case the desired covering exists.
\end{proof}
For application, assume $\tr E=q\leq 1$ for all
$E\in{\cal E}$. Then a consequence of the estimates of the lemma is
$$\|\rho-\bar{\rho}\|_1\leq (\epsilon+\tau)+\sqrt{8(\epsilon+\tau)}.$$
To see this observe
\begin{equation*}\begin{split}
  \|\rho-\bar{\rho}\|_1 &\leq \|\rho-\Pi_1\rho\Pi_1\|_1
                             +\|\Pi_1\rho\Pi_1-\Pi_1\bar{\rho}\Pi_1\|_1 \\
                        &\qquad +\|\bar{\rho}-\Pi_1\bar{\rho}\Pi_1\|_1  \\
                        &\leq \tau+\epsilon+\sqrt{8(\epsilon+\tau)},
\end{split}\end{equation*}
where the three terms are estimated as follows: for the first
note $\rho-\Pi_1\rho\Pi_1=\Pi_0\rho\Pi_0$, and apply
$\tr\rho\Pi_0\leq\tau$.
For the second use the lemma, and for the third
use $\tr\bar{\rho}\Pi_0\leq \epsilon+\tau$ in lemma~V.9
of~\cite{winter:qstrong}.
\par\bigskip
We collect here a number of standard facts about types and
typical sequences (cf.~\cite{csiszar:koerner}):
\par\noindent
\emph{Empirical distributions ({\rm aka} types)}: For a probability
distribution $P$ on $\fset{X}$ define
$$\fset{T}^n_P=\{x^n\in\fset{X}^n:\forall x\in\fset{X}\ N(x|x^n)=nP(x)\},$$
where $N(x|x^n)$ counts the number of $x$'s in $x^n$. If this set is nonempty,
we call $P$ an \emph{$n$--distribution}, or \emph{type},
or \emph{empirical distribution}. Notice that the number of types is
$$\left(\!\!\begin{array}{c}
              {n+a-1}\\ {a-1}
            \end{array}\!\!\right)\leq (n+1)^a.$$
\par\noindent
\emph{Typical sequences}: For $\alpha\geq 0$ and any distribution $P$
on $\fset{X}$ define the following set of \emph{typical sequences}:
\begin{equation*}\begin{split}
  \fset{T}^n_{P,\alpha} &=\bigg\{x^n:\forall x\ |N(x|x^n)-nP(x)| \\
                        &\phantom{==========}\leq\alpha\sqrt{n}\sqrt{P(x)(1-P(x))}\bigg\} \\
                        &=\bigcup_{Q\text{ s.t. }|Q(x)-P(x)|\leq 
                                            \sqrt{\frac{P(1-P)}{n}}} \fset{T}^n_Q
\end{split}\end{equation*}
Note that by Chebyshev inequality
$P^{\otimes n}(\fset{T}^n_{P,\alpha})\geq 1-\frac{a}{\alpha^2}$.
\par\bigskip
From~\cite{winter:qstrong} recall the following facts about the
quantum version of the previous constructions:
\par\noindent
\emph{Typical subspace}: For $\rho$ on ${\cal H}$ and $\alpha\geq 0$
there exists an orthogonal
subspace projector $\Pi^n_{\rho,\alpha}$ commuting with
$\rho^{\otimes n}$, and satisfying
\begin{align*}
  \label{eq:typ}
  \tr(\rho^{\otimes n}\Pi^n_{\rho,\alpha}) &\geq 1-\frac{d}{\alpha^2},\tag{1}\\
  \tr\Pi^n_{\rho,\alpha}                   &\leq \exp(nH(\rho)+Kd\alpha\sqrt{n}),\\
  \Pi^n_{\rho,\alpha}\rho^{\otimes n}\Pi^n_{\rho,\alpha}
                                           &\geq \exp(-nH(\rho)-Kd\alpha\sqrt{n})
                                                   \Pi^n_{\rho,\alpha}.
\end{align*}
\par\noindent
\emph{Conditional typical subspace}: For $x^n\in\fset{T}^n_P$ and $\alpha\geq 0$
there exists an orthogonal
subspace projector $\Pi^n_{W,\alpha}(x^n)$ commuting with
$W^n_{x^n}$, and satisfying
\begin{align*}
  \label{eq:con:typ}
  \tr(W^n_{x^n}\Pi^n_{W,\alpha}(x^n))     &\geq 1-\frac{ad}{\alpha^2},\tag{2}\\
  \tr\Pi^n_{W,\alpha}(x^n)                &\leq \exp(nH(W|P)+Kda\alpha\sqrt{n}),\\
  \Pi^n_{W,\alpha}(x^n)W^n_{x^n}\Pi^n_{W,\alpha}(x^n)
                                          &\leq \exp(-nH(W|P)+Kda\alpha\sqrt{n})\!\cdot\\
                                  &\phantom{=============}\cdot\!\Pi^n_{W,\alpha}(x^n),\\
  \label{eq:con:typ:2}
  \tr(W^n_{x^n}\Pi^n_{PW,\alpha\sqrt{a}}) &\geq 1-\frac{ad}{\alpha^2}.\tag{3}
\end{align*}
\par\bigskip
\emph{Proof of theorem~\ref{satz:QID:converse}}:
  We follow the strategy of the proof for
  theorem~\ref{satz:ID:converse}: consider a $(n,N,\lambda_1,\lambda_2)$
  QID code $\{(P_i,D_i):i=1,\ldots,N\}$, $\lambda_1+\lambda_2=1-\lambda<1$,
  and concentrate on one $P_i$ for the moment.
  Introduce, for empirical distributions $T$ on ${\cal X}$,
  the probability distributions
  $$P^T_i(x^n)=\frac{P_i(x^n)}{P_i({\cal T}^n_T)}\text{  for }x^n\in{\cal T}^n_T,$$
  extended by $0$ to ${\cal X}^n$.
  For $x^n\in{\cal T}^n_T$ and with
  $$\alpha=\frac{\sqrt{600ad}}{\lambda},$$
  construct the conditional
  typical projector $\Pi^n_{W,\alpha}(x^n)$,
  and the typical projector $\Pi^n_{TW,\alpha\sqrt{a}}$.
  \par
  Define the operators
  $$Q_{x^n}=\Pi^n_{TW,\alpha\sqrt{a}}\Pi^n_{W,\alpha}(x^n)
                           W^n_{x^n}
            \Pi^n_{W,\alpha}(x^n)\Pi^n_{TW,\alpha\sqrt{a}},$$
  and note that
  $$\|Q_{x^n}-W^n_{x^n}\|_1\leq\frac{\lambda}{6},$$
  by equations (\ref{eq:con:typ}) and (\ref{eq:con:typ:2}),
  and lemma~V.9 of~\cite{winter:qstrong}.
  \par
  Now we apply lemma~\ref{lemma:qu:covering} with
  $\epsilon=\tau=\lambda^2/1200$ to the quantum hypergraph with
  the range of $\Pi^n_{TW,\alpha\sqrt{a}}$ as vertex space and
  edges
  $$\Pi^n_{TW,\alpha\sqrt{a}}
      \Pi^n_{W,\alpha}(x^n)W^n_{x^n}\Pi^n_{W,\alpha}(x^n)
    \Pi^n_{TW,\alpha\sqrt{a}},\ x^n\in{\cal T}^n_T.$$
  Combining we get a $L$--distribution $\bar{P}^T_i$ with
  $$\|P^T_i Q-\bar{P}^T_i Q\|_1\leq\frac{\lambda}{6},$$
  $$L\leq \exp(nI(T;W)+O(\sqrt{n}))\leq \exp(nC(W)+O(\sqrt{n})),$$
  where the constants depend explicitly on $\alpha,\delta,\tau$.
  By construction we get
  $$\|P^T_iW^n-\bar{P}^T_iW^n\|_1\leq\frac{\lambda}{3}.$$
  By the proof of lemma~\ref{lemma:qu:covering} we can choose
  $L=\exp(nC(W)+O(\sqrt{n}))$, independent of $i$ and $T$.
  \par
  Now choose a $K$--distribution $R$ on the set of all empirical
  distributions such that
  $$\sum_{T\text{ emp. distr.}} |P_i({\cal T}^n_T)-R(T)|\leq\frac{\lambda}{3},$$
  which is possible for
  $$K=\lceil 3(n+1)^{|{\cal X}|}/\lambda \rceil.$$
  Defining then
  $$\bar{P}_i=\sum_{T\text{ emp. distr.}} R(T)\bar{P}^T_i$$
  we deduce finally
  $$\frac{1}{2}\|P_iW^n-\bar{P}_iW^n\|_1\leq\frac{\lambda}{3}.$$
  Since for every operator $D$ on ${\cal H}^{\otimes n}$,
  $0\leq D\leq \1$
  $$|\tr(P_iW^n\!\cdot\! D)-\tr(\bar{P}_iW^n\!\cdot\! D)|\leq
                               \frac{1}{2}\|P_iW^n-\bar{P}_iW^n\|_1$$
  the collection $\{(\bar{P}_i,D_i):i=1,\ldots,N\}$ is indeed a
  $(n,N,\lambda_1+\lambda/3,\lambda_2+\lambda/3)$ QID code.
  \par
  The proof is concluded by two observations: because of
  $\lambda_1+\lambda_2+2\lambda/3<1$ we have
  $\bar{P}_i\neq\bar{P}_j$ for $i\neq j$. Since the $\bar{P}_i$
  however are $KL$--distributions, we find
  \begin{equation*}\begin{split}
    N &\leq |{\cal X}^n|^{KL}=\exp(n\log|{\cal X}|\cdot{KL}) \\
      &\leq \exp(\exp(n(C(W)+\delta))),
  \end{split}\end{equation*}
  the last if only $n$ is large enough.
\qed\par
We note that we actually proved the upper bound
$C(W)$ to the resolution of a discrete memoryless quantum channel,
in the following sense:
\begin{defi}
  \label{def:resolvability}
  Let ${\bf W}$ be any quantum channel, i.e.~a family ${\bf W}=(W^1,W^2,\ldots)$
  of maps
  $$W^n:{\cal X}^n\rightarrow {\cal S}({\cal H}^{\otimes n}).$$
  A number $R$ is called $\epsilon$--\emph{achievable resolution rate}
  if for all $\delta>0$ there is $n_0$ such that for all $n\geq n_0$
  and all probability distributions $P^n$ on ${\cal X}^n$
  there is an $M$--distribution $Q^n$ on ${\cal X}^n$ with the
  properties
  $$M\leq\exp(n(R+\delta))\text{ and }\|P^nW^n-Q^nW^n\|_1\leq\epsilon.$$
  Define
  $$S_\epsilon=\inf\{R:R\text{ is }\epsilon\text{--achievable resolution rate}\},$$
  the channel's $\epsilon$--\emph{resolution}.
\end{defi}
Observe that this goes beyond the definition of~\cite{loeber:diss},
where the resolution was a function of a measurement process
${\bf E}$:
\begin{defi}[L\"ober~\cite{loeber:diss}]
  \label{def:sim:resolvability}
  Let ${\bf E}=(E^1,E^2,\ldots)$ be a sequence of POVMs $E_n$ on
  ${\cal H}^{\otimes n}$, and adopt the notations of
  definition~\ref{def:resolvability}.
  A number $R$ is called $\epsilon$--\emph{achievable resolution rate for}
  ${\bf E}$ if for all $\delta>0$ there is $n_0$ such that for all $n\geq n_0$
  and all probability distributions $P^n$ on ${\cal X}^n$
  there is an $M$--distribution $Q^n$ on ${\cal X}^n$ with the
  properties
  $$M\leq\exp(n(R+\delta))\text{ and }d_{E^n}(P^nW^n,Q^nW^n)\leq\epsilon,$$
  where $d_{E}(\rho,\sigma)$ is the total variational distance
  of the two output distributions generated by applying $E$ to states
  $\rho$, $\sigma$, respectively.
  \par\noindent
  Define
  $$S_\epsilon({\bf E})=\inf\{R\text{ is }\epsilon
                                 \text{--achievable resolution rate for }{\bf E}\},$$
  the channel's $\epsilon$--\emph{resolution for} ${\bf E}$.
\end{defi}
In general
$$S_\epsilon({\bf E})\leq S_\epsilon.$$
We do not know if there is an example of a channel such that
$$\sup_{\bf E} S_\epsilon({\bf E}) < S_\epsilon.$$

\section*{Acknowledgements}
Conversations with Alexander~S.~Holevo,
Friedrich G\"otze, and Peter Eichelsbacher about the theory of
nonreal random variables are acknowledged.
\par
The research of AW was partially supported by SFB 343 ``Diskrete
Strukturen in der Mathematik'' of the Deutsche Forschungsgemeinschaft.

\appendix
\section*{Operator valued random variables}

\subsection{Introduction}
  \label{subsec:intro}
  The theory of real random variables provides the framework of much
  of modern probability theory, such as laws of large numbers,
  limit theorems, and probability estimates for `deviations',
  when sums of independent random variables are involved.
  However several authors have started to develop analogous
  theories for the case that the algebraic structure of the reals
  is substituted by more general structures such as groups,
  vector spaces, etc., see for example~\cite{grenander:algebraic}.
  \par
  In the present work we focus on a structure that has vital
  interest in quantum probability theory, namely the algebra
  of operators on a (complex) Hilbert space, and in particular
  the real vector space of selfadjoint operators therein which
  can be regarded as a partially ordered generalization of
  the reals (as embedded in the complex numbers). In particular
  it makes sense to discuss probability estimates as the
  Markov and Chebyshev inequality (subsection C),
  and in fact one can even generalize the exponentially good
  estimates for large deviations by the so--called Bernstein
  trick which yield the famous Chernoff bounds
  (subsection D).
  \par
  Otherwise the plan of this appendix is as follows:
  subsection B
  collects basic definitions and notation
  we employ, and some facts from the theory of operator and
  trace inequalities, after the central subsections C
  and D
  we collect a number of plausible
  conjectures (subsection E),
  and close
  with an application to the noncommutative generalization
  of the covering problem for hypergraphs, in
  subsection F.

\subsection{Basic facts and definitions}
  \label{subsec:basic}
  We will study random variables $X:\Omega\longrightarrow\alg{A}_s$,
  where $\alg{A}_s=\{A\in\alg{A}:A=A^*\}$ is the selfadjoint part
  of the C${}^*$--algebra $\alg{A}$, which is a real vector space.
  Usually we will restrict our attention to the most interesting
  and in a sense generic case of the full operator algebra
  $\alg{L}({\cal H})$ of the complex Hilbert space ${\cal H}$.
  Throughout the paper we denote $d=\dim{\cal H}$, which we assume
  to be finite. In the general case $d=\tr\1$, and $\alg{A}$
  can be embedded into $\alg{L}(\C^d)$ as an algebra, preserving
  the trace.
  \par
  The real cone $\alg{A}_+=\{A\in\alg{A}:A=A^*\geq 0\}$
  induces a partial order $\leq$ in $\alg{A}_s$, which will be the
  main object of interest in what follows. Let us introduce some
  convenient notation: for $A,B\in\alg{A}_s$ the \emph{closed interval}
  $[A,B]$ is defined as
  $$[A,B]=\{X\in\alg{A}_s:A\leq X\leq B\}.$$
  (Similarly open and halfopen intervals $(A,B)$, $[A,B)$, etc.).
  \par
  For simplicity we will assume that the space $\Omega$ on which the
  random variables live is discrete.
  \par
  Some remarks on the operator order:
  \begin{description}
    \item [(A)] $\leq$ is not a total order unless $\alg{A}=\C$,
      in which case $\alg{A}_s=\R$. Thus in this case (which we will
      refer to as the \emph{classical case}) the theory
      developed below reduces to the study of real random variables.
    \item [(B)] $A\geq 0$ is equivalent to saying that all eigenvalues
      of $A$ are nonnegative. These are $d$ nonlinear inequalities.
      However from the alternative characterization
      \begin{align*}
        A\geq 0 &\Longleftrightarrow
                   \forall\rho\text{ density operator }\tr(\rho A)\geq 0 \\
                &\Longleftrightarrow
                   \forall\pi\text{ one--dim. projector }\tr(\pi A)\geq 0
      \end{align*}
      we see that this is equivalent to infinitely many \emph{linear}
      inequalities, which is better adapted to the vector space structure
      of $\alg{A}_s$.
    \item [(C)] The operator mappings $A\mapsto A^s$ (for $s\in[0,1]$)
      and $A\mapsto\log A$ are defined on $\alg{A}_+$, and both are
      operator monotone and operator concave.\\
      In contrast, the mappings $A\mapsto A^s$ (for $s>2$) and
      $A\mapsto\exp A$ are neither operator monotone nor operator convex.
      Interestingly, $A^s$ for $s\in[1,2]$ \emph{is} operator convex
      (though not operator monotone).\\
      All this follows from L\"owner's theorem~\cite{loewner:thm}, a good account
      of which is given in Donoghue's book~\cite{donoghue:monotone}, and
      a characterization of operator convex functions due to
      Hansen and Pedersen~\cite{hansen:pedersen}.
    \item [(D)] Note however that the mapping $A\mapsto\tr\exp A$ \emph{is}
      monotone and convex: see Lieb~\cite{lieb:WYD}.
    \item [(E)] Golden--Thompson--inequality
      (\cite{golden:inequality,thompson:inequality}): for $A,B\in\alg{A}_s$
      $$\tr\exp(A+B)\leq\tr\left((\exp A)(\exp B)\right).$$
  \end{description}

\subsection{Markov and Chebyshev inequality}
  \label{subsec:markov}
  \begin{satz}[Markov inequality]
    \label{satz:markov}
    Let $X$ a random variable with values in
    $\alg{A}_+$ and expectation $M=\E X=\sum_{x} \Pr\{X=x\}x$,
    and $A\geq 0$ (i.e.~$A\in\alg{L}({\cal H})_+$).
    Then
    $$Pr\{X\not\leq A\}\leq \tr\left(MA^{-1}\right).$$
  \end{satz}
  \begin{proof}
    We may assume that the support of $A$ contains the support of $M$,
    otherwise the theorem is trivial.\par
    Consider the positive random variable
    $$Y=A^{-1/2}XA^{-1/2},$$
    which has expectation $\E Y=N=A^{-1/2}MA^{-1/2}$. Since the events
    $\{X\leq A\}$ and $\{Y\leq\1\}$ coincide we have to show that
    $$Pr\{Y\not\leq\1\}\leq\tr N.$$
    This is seen as follows:
    $$N=\sum_y \Pr\{Y=y\}y\geq \sum_{y\not\leq\1} \Pr\{Y=y\}y.$$
    Taking traces, and observing that a positive operator which
    is not less than or equal $\1$ must have trace at least 1, we
    find
    \begin{equation*}\begin{split}
      \tr N &\geq \sum_{y\not\leq\1} \Pr\{Y=y\}\tr y \\
            &\geq \sum_{y\not\leq\1} \Pr\{Y=y\} = \Pr\{Y\not\leq\1\},
    \end{split}\end{equation*}
    which is what we wanted.
  \end{proof}
  \begin{bem}
    In the case of ${\cal H}=\C$ the theorem reduces to the well known
    Markov inequality for nonnegative real random variables. One can easily
    see that like in this classical case the inequality of the theorem
    is optimal in the sense that there are examples when it is assumed
    with equality.
  \end{bem}
  If we assume knowledge about the second moment of $X$ we can prove
  \begin{satz}[Chebyshev inequality]
    \label{satz:chebyshev}
    Let $X$ a random variable with values in $\alg{A}_s$,
    expectation $M=\E X$, and variance
    $\Var X=S^2=\E\left((X-M)^2\right)=\E(X^2)-M^2$.
    For $\Delta\geq 0$
    $$\Pr\{|X-M|\not\leq\Delta\}\leq\tr\left(S^2\Delta^{-2}\right).$$
  \end{satz}
  \begin{proof}
    Observing
    $$|X-M|\leq\Delta\Longleftarrow(X-M)^2\leq\Delta^2$$
    (because $\sqrt{\phantom{x}}$ is operator monotone, see
    section B,
    \emph{(C)}) we find
    \begin{align*}
      \Pr\{|X-M|\not\leq\Delta\} &\leq\Pr\{(X-M)^2\not\leq\Delta^2\} \\
                                 &\leq\tr\left(S^2\Delta^{-2}\right)
    \end{align*}
    (by theorem~\ref{satz:markov}).
  \end{proof}
  \begin{bem}
    If $X,Y$ are independent, then $\Var(X+Y)=\Var X+\Var Y$. The calculation
    is the same as in the classical case, but one has to take care of the
    noncommutativity.
  \end{bem}
  \begin{cor}[Weak law of large numbers]
    \label{cor:weak:law}
    Let $X$, $X_1$, $\ldots, X_n$ i.i.d.~random variables with $\E X=M$,
    $\Var X=S^2$, and $\Delta\geq 0$. Then
    $$\Pr\left\{\frac{1}{n}\sum_{i=1}^{n} X_i\not\in[M-\Delta,M+\Delta]\right\}
                   \leq \frac{1}{n}\tr\left(S^2\Delta^{-2}\right),$$
    $$\Pr\left\{\sum_{i=1}^{n} X_i\not\in[nM-\Delta\sqrt{n},nM+\Delta\sqrt{n}]\right\}
                   \leq \tr\left(S^2\Delta^{-2}\right).$$
  \end{cor}
  \begin{proof}
    Observe that $Y\not\in[M-\Delta,M+\Delta]$ is equivalent to
    $|Y-M|\not\leq\Delta$, and apply the previous theorem.
  \end{proof}

\subsection{Large deviations and Bernstein trick}
  \label{subsec:bernstein}
  \begin{lemma}
    \label{lemma:large:dev}
    For a random variable $Y$, $B\in\alg{A}_s$, and $T\in\alg{A}$
    such that $T^*T>0$
    $$\Pr\{Y\not\leq B\}\leq\tr\left(\E\exp(TYT^*-TBT^*)\right).$$
  \end{lemma}
  \begin{proof}
    A direct calculation:
    \begin{align*}
      \Pr\{Y\not\leq B\} &=\Pr\{Y-B\not\leq 0\} \\
                         &=\Pr\{TYT^*-TBT^*\not\leq 0\} \\
                         &=\Pr\{\exp(TYT^*-TBT^*)\not\leq\1\} \\
                         &\leq\tr\left(\E\exp(TYT^*-TBT^*)\right).
    \end{align*}
    Here the second line is because the mapping $X\mapsto TXT^*$ is bijective
    and preserves the order, the third because for \emph{commuting} operators
    $A,B$, $A\leq B$ is equivalent to $\exp A\leq\exp B$, and the last line
    by theorem~\ref{satz:markov}.
  \end{proof}
  \begin{satz}
    \label{satz:large:dev}
    Let $X,X_1,\ldots,X_n$ i.i.d.~random variables with values in $\alg{A}_s$,
    $A\in\alg{A}_s$. Then for $T\in\alg{A}$, $T^*T>0$
    $$\Pr\left\{\sum_{i=1}^{n}X_i\not\leq nA\right\}
            \leq d\cdot\left\|\E\exp\left(TXT^*-TAT^*\right)\right\|^n.$$
  \end{satz}
  \begin{proof}
    Using the previous theorem with $Y=\sum_{i=1}^n X_i$ and $B=nA$
    we find 
    \begin{align*}
      \Pr\left\{\sum_{i=1}^{n}X_i\not\leq nA\right\}
                &\leq \tr\left(\E\exp\left(\sum_{i=1}^n T(X_i-A)T^*\right)\right)   \\
                &=    \E\tr\exp\left(\sum_{i=1}^n T(X_i-A)T^*\right)                \\
                &\leq \E\tr\left[\exp\left(\sum_{i=1}^{n-1} T(X_i-A)T^*\right)\right. \\
                &\phantom{========}\cdot\exp\left(T(X_n-A)T^*\right)\Bigg]             \\
                &=\E_{1\ldots n-1}\tr\!\!\left[\exp\left(\sum_{i=1}^{n-1}
                                                   T(X_i-A)T^*\right)\right.   \\
                &\phantom{========}\cdot\E\exp\left(T(X_n-A)T^*\right)\Bigg]           \\
                &\leq \|\E\exp\left(T(X_n-A)T^*\right)\|\cdot                       \\
                &\phantom{=}\cdot\E_{1\ldots n-1}\tr\exp\left(\sum_{i=1}^{n-1}
                                    T(X_i-A)T^*\right)                              \\
                &\leq\ldots\leq d\cdot\|\E\exp\left(T(X_n-A)T^*\right)\|^n.
    \end{align*}
    Here everything is straightforward, except for the third line which
    is by the Golden--Thompson--inequality (section B,
    \emph{(E)}).
  \end{proof}
  The problem is now to minimize $\left\|\E\exp\left(TXT^*-TAT^*\right)\right\|$
  with respect to $T$. Observe that without loss of generality we may
  assume that $T$ is selfadjoint, because of the polar decomposition
  $T=U\cdot|T|$, with a unitary $U$. The case we will pursue further
  is that of a bounded random variable. Introducing the binary I--divergence
  $$D(u\|v)=u(\log u-\log v)+(1-u)\left(\log(1-u)-\log(1-v)\right)$$
  we find
  \begin{satz}[Chernoff]
    \label{satz:chernoff}
    Let $X,X_1,\ldots,X_n$ i.i.d.~random variables with values in
    $[0,\1]\subset\alg{A}_s$, $\E X\leq m\1$, $A\geq a\1$,
    $1\geq a\geq m\geq 0$. Then
    $$\Pr\left\{\sum_{i=1}^{n}X_i\not\leq nA\right\}
                   \leq d\cdot\exp\left(-nD(a\|m)\right).$$
    Similarly, if $\E X\geq m\1$, $A\leq a\1$,
    $0\leq a\leq m\leq 1$. Then
    $$\Pr\left\{\sum_{i=1}^{n}X_i\not\geq nA\right\}
                   \leq d\cdot\exp\left(-nD(a\|m)\right).$$
    As a consequence we get, for
    $\E X=M\geq \mu\1$ and $0\leq\epsilon\leq \frac{1}{2}$,
    \begin{equation*}\begin{split}
      \Pr\left\{\frac{1}{n}\sum_{i=1}^n X_i\not\in[(1-\epsilon)M,(1+\epsilon)M]\right\}& \\
      &\!\!\!\!\!\!\!\!\!\!\leq 2d\!\cdot\!\exp\left(-n\!\cdot\! \frac{\epsilon^2\mu}{2\ln 2}\right)\! .
    \end{split}\end{equation*}
  \end{satz}
  \begin{proof}
    The second part follows from the first by considering
    $Y_i=\1-X_i$, and the observation that $D(a\|m)=D(1-a\|1-m)$.\par
    To prove it we apply theorem~\ref{satz:large:dev} with
    $T=\sqrt{t}\1$:
    \begin{align*}
      \Pr\left\{\sum_{i=1}^{n}X_i\not\leq nA\right\}
                &\leq \Pr\left\{\sum_{i=1}^{n}X_i\not\leq na\1\right\} \\
                &\leq d\cdot\left\|\E\exp(tX)\exp(-ta)\right\|^n.
    \end{align*}
    Now using
    $$\exp(tX)-\1\leq X(\exp(t)-1)$$
    (which follows from the validity of the estimate for real $x$,
    $x\in(0,1)$:
    $$\frac{\exp(tx)-1}{x}\leq\frac{\exp(t)-1}{1},$$
    which in turn is just the convexity of $\exp$) we find
    \begin{align*}
      \E\exp(tX) &\leq\1+\E X(\exp(t)-1) \\
                 &\leq(1-m+m\exp t)\1.
    \end{align*}
    Hence
    $$\left\|\E\exp(tX)\exp(-ta)\right\|\leq(1-m+m\exp t)\exp(-at),$$
    and choosing
    $$t=\log\left(\frac{a}{m}\cdot\frac{1-m}{1-a}\right)>0$$
    the right hand side becomes exactly $\exp\left(-D(a\|m)\right)$.
    \par
    To prove the last claim of the theorem consider the variables
    $Y_i=\mu M^{-1/2}X_iM^{-1/2}$ with expectation
    $\E Y_i=\mu$ and $Y_i\in [0,\1]$, by hypotheses. Because of
    \begin{equation*}\begin{split}
      \frac{1}{n}\sum_{i=1}^n X_i\in & [(1-\epsilon)M,(1+\epsilon)M]        \\
                                     &\qquad \Longleftrightarrow
                                       \frac{1}{n}\sum_{i=1}^n Y_i\in
                                         [(1-\epsilon)\mu\1,(1-\epsilon)\mu\1]
    \end{split}\end{equation*}
    we can apply what we just proved to obtain
    \begin{equation*}\begin{split}
      \Pr&\left\{ \frac{1}{n}\sum_{i=1}^n X_i\not\in [(1-\epsilon)M,(1+\epsilon)M] \right\} \\
         &\ \leq d\bigl[ \exp(-nD( (1-\epsilon)\mu\|\mu ))
                            +\exp(-nD( (1+\epsilon)\mu\|\mu )) \bigr]                       \\
         &\ \leq 2d\!\cdot\!\exp\left(-n\!\cdot\! \frac{\epsilon^2\mu}{2\ln 2}\right)\!,
    \end{split}\end{equation*}
    the last line by the already used inequality
    $D((1+x)\mu\|\mu )\geq \frac{1}{2\ln 2}x^2\mu$.
  \end{proof}

\subsection{Conjectures}
  \label{subsec:conjectures}
  We have the feeling that in the estimates of the previous section
  we waste too much. In particular the theorems become useless in the
  infinite dimensional case, because in the traces we could only
  account for the supremum of the involved eigenvalues, multiplied
  by the dimension of the underlying space.
  \begin{conj}
    \label{conj:large:dev}
    Under the assumptions of theorem~\ref{satz:large:dev} it
    even holds that
    $$\Pr\left\{\sum_{i=1}^{n}X_i\not\leq nA\right\}
            \leq\tr\left[\left(\E\exp\left(TXT^*-TAT^*\right)\right)^n\right],$$
    since we conjecture that for i.i.d.~random variables
    $Z$, $Z_1,\ldots,Z_n\in\alg{A}_s$
    $$\tr\E\exp\left(\sum_{i=1}^n Z_i\right)\leq\tr\left(\left(\E\exp(Z)\right)^n\right).$$
  \end{conj}
  Note that this is indeed true for $n=2$, thanks to the Golden--Thompson
  inequality!
  \par
  For larger $n$ there seems to be no applicable generalization of the
  Golden--Thompson inequality, so a different approach is
  needed. We propose to take logarithms in the above conjecture instead 
  of traces: by the monotonicity of $\tr\exp A$ the conjecture is true
  if
  $$\log\E\exp\left(\sum_{i=1}^n Z_i\right)\leq n\log\E\exp(Z).$$
  Thus by induction and monotonicity of $\tr\exp A$ we can indeed prove
  conjecture~\ref{conj:large:dev} if the following is true:
  \begin{conj}
    \label{conj:log:exp}
    For finite families of selfadjoint operators $A_i$ and $B_j$
    \begin{equation*}\begin{split}
      \log &\left(\sum_{ij}\exp(A_i+B_j)\right)\leq \\
           &\phantom{\log\log\log}\log\left(\sum_{i}\exp A_i\right)
                                 +\log\left(\sum_{j}\exp B_j\right).
    \end{split}\end{equation*}
    Note that if all $A_i,B_j$ commute then equality holds!
  \end{conj}
  It may be that taking this for granted one can prove the following
  conjecture (compare with theorem~\ref{satz:chernoff}):
  \begin{conj}
    \label{conj:chernoff}
    Let $X,X_1,\ldots,X_n$ i.i.d.~random variables with values in
    $[0,\1]$, $\E X\leq M\leq A\leq\1$. Then
    $$\Pr\left\{\sum_{i=1}^{n}X_i\not\leq nA\right\}
            \leq\tr\exp\left(-n{\cal D}(A\|M)\right),$$
    where ${\cal D}$ is the operator version of the binary
    I--divergence:
    \begin{align*}
      {\cal D}(A\|M) &=\sqrt{A}(\log A-\log M)\sqrt{A} \\
                     &\phantom{=}+\sqrt{\1-A}\left(\log(\1-A)-\log(\1-M)\right)\sqrt{\1-A}.
    \end{align*}
  \end{conj}
  Again, given conjecture~\ref{conj:log:exp}, it would suffice to compare
  $\log\E\exp(TXT^*-TAT^*)$ and ${\cal D}(A\|M)$, for a clever choice
  of $T$.

\subsection{An application}
  \label{subsec:applications}
  In this last subsection we want to discuss one application of our
  estimates in ``noncommutative combinatorics'', namely as a tool
  in applying the probabilistic method to the noncommutative
  analogue of covering hypergraphs. Apart from the application
  in the main text, we would like to point out
  two other ones: an approximation
  problem in quantum estimation theory~\cite{massar:winter:compress},
  and its generalization to asymptotic convex decompositions of
  POVMs~\cite{winter:data-in-qm}.

  \subsubsection{Noncommutative hypergraphs}
    \label{subsubsec:hypergraphs}
    We will define noncommutative hypergraphs as generalizations
    of the usual ones. To understand the following definition one
    has to recall the correspondence between a compact space $X$ and
    the C${}^*$--algebra $C(X)$ of its continuous $\C$--valued functions,
    provided by the Gelfand--Naimark theorem
    (see~\cite{bratteli:robinson:1}, ch.~2.3). In the case
    of a finite discrete set this is summarized in the fact that
    the positive idempotents of the function algebra are exactly
    the characteristic functions of subsets. Thus we can talk about
    hypergraphs $({\cal V},{\cal E})$ --- ${\cal V}$ is the finite
    \emph{vertex} set, and ${\cal E}\subset 2^{\cal V}$ the set
    of \emph{hyperedges} (or \emph{edges} for short) --- in the
    language of finite dimensional commutative C${}^*$--algebras
    and certain of their idempotents.
    \par
    A \emph{noncommutative hypergraph} $\Gamma$ is a pair $(\alg{V},{\cal E})$
    with a finite dimensional C${}^*$--algebra $\alg{V}$ and a set
    ${\cal E}\subset[0,\1]$ (usually finite). We call $\Gamma$ \emph{strict}
    if all elements of ${\cal E}$ are idempotents. Finally
    $\Gamma$ is a \emph{quantum hypergraph} if $\alg{V}$ is the full operator
    algebra of a finite dimensional complex Hilbert spave ${\cal V}$, in
    which case we denote $\Gamma$ as $({\cal V},{\cal E})$.
    From the theory of finite dimensional C${}^*$--algebras it is known that
    $\alg{V}$ can be embedded into the full operator algebra of a
    Hilbert space of dimension $\tr\1$, preserving the trace. Thus we will
    in the sequel always assume that we deal with quantum hypergraphs.
    \par
    For a finite edge set ${\cal E}$ the \emph{degree} is defined as
    the operator
    $$\deg_{\cal E}=\sum_{E\in{\cal E}} E.$$
    A \emph{covering} of $\Gamma=({\cal V},{\cal E})$ is a finite
    family ${\cal C}$ of edges such that $\deg_{\cal C}\geq\1$.

  \subsubsection{Covering theorems}
    \label{subsubsec:covering}
    Now we come to our first covering theorem (for the classical case
    compare~\cite{ahlswede:cov:col:1}):
    \begin{satz}
      \label{satz:covering:1}
      Let $\Gamma$ a quantum hypergraph with
      $$\deg_\Gamma\geq\delta\1.$$
      Then there exists a covering of $\Gamma$ with
      $k\leq 1+\frac{8|{\cal E}|\ln 2}{\delta}\log d$ many edges.
    \end{satz}
    This is the special case of the uniform distribution in the following
    \begin{satz}
      \label{satz:covering:2}
      Let $\Gamma$ a quantum hypergraph and $P$ a probability distribution
      on ${\cal E}$, such that
      $$\sum_{E\in{\cal E}} P(E)E\geq\mu\1.$$
      Then there exists a covering of $\Gamma$ with
      $k\leq 1+8(\ln 2\log d)\mu^{-1}$ many edges.
    \end{satz}
    \begin{proof}
      Draw edges at random, i.e.~consider i.i.d.~random variables
      $X,X_1,\ldots,X_k$ with $\Pr\{X=E\}=P(E)$.
      Then we obtain, using theorem~\ref{satz:chernoff}:
      \begin{align*}
        \Pr\left\{\sum_{i=1}^k X_i\not\geq\1\right\}
                      &=\Pr\left\{\sum_{i=1}^k X_i\not\geq k\cdot\frac{1}{k}\1\right\}  \\
                      &\leq d\exp\left(-k\cdot D\left(\frac{1}{k}\|\mu\right)\right)    \\
                      &\leq d\exp\left(-k\cdot D\left(\frac{1}{2}\mu\|\mu\right)\right) \\
                      &\leq d\exp\left(-k\cdot \frac{1}{8\ln 2}\mu\right),
      \end{align*}
      the third line only if we have $k\geq 2\mu^{-1}$. Now the
      last expression is smaller than $1$ for
      $$k>8(\ln 2\log d)\mu^{-1},$$
      justifying ex post our estimates. Hence for $k$ as in the
      theorem there exists a covering with $k$ edges.
    \end{proof}
    We apply this result to a generalization of a result on covering
    numbers of hypergraphs, due to Posner and McEliece~\cite{posner:mceliece},
    obtained independently, but a little bit later,
    by Ahlswede and reported in~\cite{ahlswede:product:covering}.
    For a quantum hypergraph $\Gamma=({\cal V},{\cal E})$ define
    $\Gamma^n=({\cal V}^{\otimes n},{\cal E}^n)$, with
    $${\cal E}^n=\{E^n=E_1\otimes\cdots\otimes E_n: E_1,\ldots,E_n\in{\cal E}\}.$$
    We are interested in the covering number
    $c(n)$ of $\Gamma^n$, i.e.~the minimum cardinality of a covering
    of $\Gamma^n$. Finally define
    $$\tilde{c}(n)=\min\left\{\sum_{E^n} v(E^n):
                                    v\geq 0,\ \sum_{E^n} v(E^n)E^n\geq\1^{\otimes n}\right\}.$$
    (The $v$ can be seen as a continous weight version of coverings, and
    will be called \emph{generalized coverings}).
    It is immediate that $c(n)\geq\tilde{c}(n)$.
    \begin{satz}
      \label{satz:covering:products}
      With
      $$C=-\log\left(\max_{P\text{ p.d. on }{\cal E}}
                         \min\sum_{E\in{\cal E}} P(E)E\right),$$
      where the $\min$ means the minimal eigenvalue, one has
      \begin{align*}
        \tilde{c}(n) &\geq\exp(Cn), \\
        c(n)         &\leq 1+8(\ln 2\log d)\cdot n\exp(Cn).
      \end{align*}
      In particular
      $$\lim_{n\rightarrow\infty} \frac{1}{n}\log c(n)
         =\lim_{n\rightarrow\infty} \frac{1}{n}\log\tilde{c}(n)
         =C.$$
    \end{satz}
    \begin{proof}
      The second estimate follows by applying theorem~\ref{satz:covering:2}
      with the distribution $P^{\otimes n}$.
      \par
      The first is proved by induction on $n$. The case $n=0$ is trivial,
      and the case $n=1$ is seen as follows:
      let $v^*$ be a minimal weight generalized covering of $\Gamma$, i.e.
      $$\tilde{c}(1)=\sum_E v^*(E)\text{ and } \sum_E V^*(E)E\geq \1.$$
      What we have to show is that
      $$\tilde{c}(1)\geq\left(\max_P\min\sum_E P(E)E\right)^{-1},$$
      which means we have to find a distribution $P$ such that
      $$sum_E P(E)E\geq {\tilde{c}(1)}^{-1}\1.$$
      With $P(E)=v^*(E){\tilde{c}(1)}^{-1}$ this is obviously satisfied.
      \par
      Now assume $n>0$, and let $v^*$ a minimal weight generalized
      covering of $\Gamma^n$. Define a probability distribution $Q$
      on ${\cal E}$ by
      $$Q(E)=\frac{1}{\tilde{c}(n)}\sum_{E^n\in{\cal E}^n,E_n=E} v^*(E^n).$$
      Multiplying the relation
      $$\sum_{E^n} v^*(E^n)E^n\geq\1^{\otimes n}$$
      by $\1^{\otimes(n-1)}\otimes\pi$ (for a one--dimensional projector
      $\pi$ on ${\cal V}$) from both sides and taking
      the trace over the last factor we find
      \begin{align*}
        \1^{\otimes(n-1)} &\leq\sum_{E_n\in{\cal E}}
                              \tr(\pi E_n)\sum_{E^{n-1}\in{\cal E}^{n-1}}
                                                            v^*(E^n)E^{n-1} \\
                          &=\sum_{E^{n-1}\in{\cal E}^{n-1}}\left(
                              \sum_{E_n\in{\cal E}}\tr(\pi E_n)v^*(E^n)\right)E^{n-1}.
      \end{align*}
      This means that we have a generalized covering of $\Gamma^{n-1}$ 
      and hence
      \begin{align*}
        \tilde{c}(n-1) &\leq \sum_{E_n\in{\cal E}} \tr(\pi E_n)
                               \sum_{E^{n-1}\in{\cal E}^{n-1}} v^*(E^n)\\
                       &=\sum_{E_n\in{\cal E}} \tr(\pi E_n)Q(E_n)\tilde{c}(n).
      \end{align*}
      Thus for all $\pi$
      $$\sum_{E\in{\cal E}} \tr(\pi E)Q(E)\geq\frac{\tilde{c}(n-1)}{\tilde{c}(n)},$$
      which implies
      $$\min\sum_{E\in{\cal E}} Q(E)E\geq\frac{\tilde{c}(n-1)}{\tilde{c}(n)},$$
      which in turn implies
      $$\exp(-C)=\max_P\min\sum_{E\in{\cal E}} P(E)E\geq
                                         \frac{\tilde{c}(n-1)}{\tilde{c}(n)}.$$
    \end{proof}

\nocite{*}
\bibliographystyle{IEEE}

\begin{thebibliography}{M}
  \bibitem{ahlswede:cov:col:1} R. Ahlswede,
    ``Coloring {H}ypergraphs: {A} {N}ew {A}pproach to {M}ulti--user
    {S}ource {C}oding --- {I}'',
    J. Combinatorics, Information \& System Sciences, vol. 4, no.
    1, pp. 76--115, 1979.

  \bibitem{ahlswede:product:covering} R. Ahlswede,
    ``On set coverings in cartesian product spaces'',
    Preprint E92--005, Sonderforschungsbereich 343 ``Diskrete Strukturen
    in der Mathematik'', Universit{\"{a}}t Bielefeld, 1992.

  \bibitem{ahlswede:IDconverse} R. Ahlswede, ``On concepts of performance parameters
    for channels'', to appear in IEEE Trans. Inf. Theory, Special issue in
    memory of A. D. Wyner.

  \bibitem{ahlswede:dueck:1} R. Ahlswede, G. Dueck, ``Identification via channels'',
    IEEE Trans. Inf. Theory, vol. 35, pp. 15--29, 1989.



  \bibitem{bratteli:robinson:1} O. Bratteli, D.~W. Robinson,
    \emph{Operator Algebras and Quantum Statistical Mechanics I},
    Springer--Verlag, 1979.

  \bibitem{csiszar:koerner} I. Csisz\'{a}r, J. K\"orner, \emph{Information Theory:
      Coding Theorems for Discrete Memoryless Systems}, Academic Press, New York, 1981.

  \bibitem{davies:etc} E. B. Davies, \emph{Quantum Theory of Open Systems},
    Academic Press, London, 1976.

  \bibitem{donoghue:monotone} W.~F. Donoghue, Jr.,
    \emph{Monotone Matrix Functions and Analytic Continuation}, Springer, 1974.

  \bibitem{golden:inequality} S.~Golden, ``Lower {B}ounds for the {H}elmholtz {F}unction'',
    Physical Review, vol. 137B, no. 4, pp. B1127--1128, 1965.

  \bibitem{grenander:algebraic} U. Grenander, \emph{Probabilities on Algebraic Structures},
    Wiley \& Sons, New York, London, 1963.

  \bibitem{han:verdu:1} T. S. Han, S. Verd\'{u}, ``New results in the theory of
    identification via channels'', IEEE Trans. Inf. Theory, vol. 38, no. 1, pp. 14--25, 1992.

  \bibitem{han:verdu:2} T. S. Han, S. Verd\'{u}, ``Approximation theory of
    output statistics'', IEEE Trans. Inf. Theory, vol. 39, no. 3, pp. 752--772, 1993.

  \bibitem{hansen:pedersen} F. Hansen, G. K. Pedersen, ``Jensen's Inequality
    for Operators and L\"owner's Theorem'', Math. Ann., vol. 258, pp. 229--241,
    1982.

  \bibitem{holevo:bound} A. S. Holevo, Problemy Peredachi Informatsii, vol. 9,
    no. 3, pp. 3--11, 1973 (english translation: ``Bounds for the quantity of
    information transmitted by a quantum channel'', Probl. Inf. Transm., vol. 9,
    no. 3, pp. 177--183, 1973).

  \bibitem{holevo:channels} A. S. Holevo, ``Problems in the mathematical theory
    of quantum communication channels'', Rep. Math. Phys.,
    vol. 12, no. 2, pp. 273--278, 1977. 


  \bibitem{holevo:qucapacity} A. S. Holevo, ``The Capacity of the Quantum Channel with
    General Signal States'', IEEE Trans. Inf. Theory, vol. 44, no. 1, pp. 269--273, 1998.

  \bibitem{lieb:WYD} E.~H. Lieb, ``Convex trace functions and the
    {W}igner--{Y}anase--{D}yson conjecture'',
    Adv. Math., vol. 11, pp. 267--288, 1973.

  \bibitem{loeber:diss} P. L\"ober, \emph{Quantum Channels and Simultaneous ID Coding},
    doctoral dissertation, Universit\"at Bielefeld, 1999. WWW at
    {\tt http://archiv.ub.uni-bielefeld.de/disshabi/mathe.htm}.

  \bibitem{loewner:thm} K.~L\"owner, ``{\"U}ber monotone {M}atrixfunktionen'',
    Math. Z., vol. 38, pp. 177--216, 1934.

  \bibitem{ogawa:nagaoka} T. Ogawa, H. Nagaoka, ``Strong Converse to the Quantum
    Channel Coding Theorem'', IEEE Trans. Inform. Theory,
    vol. 45, no. 7, pp. 2486--2489, 1999.

  \bibitem{posner:mceliece} R.~J. McEliece, E.~C. Posner,
    ``Hide and seek, data storage and entropy'',
    Annals Math. Statistics, vol. 42, pp. 1706--1716, 1971.

  \bibitem{schumacher:qucoding} B. Schumacher, ``Quantum Coding'',
    Phys. Rev. A, vol. 51, no. 4, pp. 2738--2747, 1995.


  \bibitem{shannon:info} C. E. Shannon, ``A mathematical theory of communication'',
    Bell System Tech. J., vol. 27, pp. 379--423; \emph{ibid.} pp. 623--656, 1948.

  \bibitem{thompson:inequality} C.~J. Thompson, ``Inequality with
    {A}pplications in {S}tatistical {M}echanics'',
    J. Math. Phys., vol. 6, no. 11, pp. 1812--1823, 1965.


  \bibitem{winter:qstrong} A. Winter, ``Coding theorem and strong converse for quantum
    channels'', IEEE Trans. Inform. Theory, 
    vol. 45, no. 7, pp. 2481--2485, 1999.

%

  \bibitem{winter:data-in-qm} A. Winter, ``Extrinsic and intrinsic data in quantum measurements:
    asymptotic convex decomposition of positive operator valued measures'', e--print
    {\tt quant-ph/0109050} at {\tt http://arXiv.org}, 2001.

  \bibitem{massar:winter:compress} A. Winter, S. Massar,
    ``Compression of quantum measurement operations'', Phys. Rev. A,
    vol. 64, 012311, 2001.

  \bibitem{wolfowitz:strong} J. Wolfowitz, ``The coding of messages subject to chance
    errors'', Illinois J. Math., vol. 1, no. 4, pp. 591--606, 1957.


\end{thebibliography}

\end{document}